\documentclass{JHEP3} 

\usepackage{epsfig,multicol,bbm}

\newcommand\fverb{\setbox\fverbbox=\hbox\bgroup\verb}
\newcommand\fverbdo{\egroup\medskip\noindent%
			\fbox{\unhbox\fverbbox}\ }
\newcommand\fverbit{\egroup\item[\fbox{\unhbox\fverbbox}]}
\newbox\fverbbox

\title{Higher derivative terms including the Ramond-Ramond five-form.}

\author{Miguel F. Paulos\\
	DAMTP, University of Cambridge\\
	E-mail: \email{M.F.Paulos@damtp.cam.ac.uk}
	}

\preprint{}

\abstract{
Superfield methods can be used to determine the precise way the self-dual five-form couples to the metric in the first non-trivial $\alpha'$ corrections to type IIB supergravity. We explicitly compute the exact tensor structure of these terms. This requires extensive use of computing algorithms to reduce the complicated expressions that appear to a surprisingly simple form. Along the way we show a new method of computing Schouten identities. 
With this result we clarify under which conditions one may neglect the five-form higher derivative terms. We comment on corrections to the thermodynamics of charged black holes.
}

\usepackage{mqcommands}

\begin{document}
\maketitle 
\section{Introduction}

The low momentum expansion of superstring theory is known to lead to an effective action for the massless mode which is simply supergravity. Going beyond lowest order in the string length leads to an effective expansion of the action in power of $\alpha'$. Higher powers of $\alpha'$ correspond to higher derivative terms which provide information on various intrinsically stringy effects, and knowing them leads to a number of applications.

These higher order corrections are relevant to black hole physics in several ways. For example, they are important for stretching the horizon if the classical black hole solution does not have one \cite{Dabholkar:2004dq}.  They also lead to a modification of the Bekenstein-Hawking area law for the entropy \cite{Wald}. Consistency of string theory demands an agreement between calculations of entropy in this setting with microscopic state counting, and therefore higher derivative corrections allows us to better understand this correspondence
 
In the context of AdS/CFT, $\alpha'$ corresponds to $1/\sqrt{\lambda}=1/\sqrt{2 g_{YM}^2 N}$, providing valuable information on gauge theory at strong coupling \cite{Gubser}\cite{Green}. They also give finite (large) coupling corrections to the infinite coupling limit of correlation functions.

We are here concerned with the corrections at first non-trivial order in the low momentum expansion of type IIB superstring theory, which are order $\alpha'^3$ with respect to the original classical supergravity action. Computing these corrections turns out to be very non-trivial, and several alternative methods have been proposed (\cite{Peeters:2003},\cite{Skenderis} and references therein).  The greatest progress has been made by calculating string scattering amplitudes and writing down an action which reproduces them. The well known $R^4$ term was calculated in this way \cite{GrossWitten}. When scattering amplitudes involve Ramond-Ramond fields the calculation is much more involved, but nevertheless some progress has been made \cite{RRsector}\cite{Policastro:2006vt}.


The order ${\alpha'}^3$ corrections may be found at the linearized level  by computing the integral of a scalar function of the linearized scalar superfield \cite{Howe:1983sra} over half the type IIB superspace \cite{Green:1999}.  Unfortunately, there are profound difficulties in formulating a supersymmetric description of the full nonlinear theory at this order \cite{Skenderis}.  However, there is a suggested exact action at order ${\alpha'}^3$ in the special case in which only the metric and five-form field strength are non-trivial [2], which was important for showing that the classical D3-brane geometry is unrenormalized at this order.  In this case it has been argued \cite{Rajaraman} that the obstruction to a chiral measure in the type IIB theory is circumvented.


In this paper the tensor structure of these corrections is explicitly computed and reduced to a manageable form. As a first step, the fermionic integral is reduced to a sum of Lorentz scalars following \cite{Superfield}. This leads to a large sum of complicated contractions of four powers of a certain tensor $\mathcal R$. It is shown that this sum must vanish unless the representation content of $\mathcal R$ is equal to $\mbf{770}\oplus \mbf{1050^+}$, so that the sum acts like a form of Young projection. Accordingly, one can write the whole set of terms as the Young-projected version of a greatly reduced set. In the process one must find basis of tensor monomials independent with respect to all the symmetries of the tensors that compose them. Along the way one finds that there are certain dimensionally-dependent identities that can be used to further reduce the number of terms. A new method for discovering these identities is also presented.

The outline is as follows. Section 2 discusses the form of the higher derivative corrections for the type IIB $\mathcal N=2$ supergravity that only involve the five-form <and the metric. The ansatz of \cite{Green} for the form of these corrections is reviewed, as well as some of its consequences for supersymmetric solutions. Section 3 is concerned with the computation that was performed to obtain these terms. A new method for discovering dimensionally dependent (Schouten) identities is described. For clarity, our results are summarized in section 4. Some applications are discussed in section 5. We comment on the conditions under which the five-form corrections might be neglected, and in particular show that the results of \cite{Gubser} are valid. Finally we perform an application of our result to the computation of $\alpha'$ corrections to the thermodynamics of charged black holes, following an approach partially justified by \cite{Gubser}. 

\section{Higher derivative corrections}
\subsection{Supersymmetric completion of $R^4$}
In this section and the following we review some basic results on higher derivative corrections to the supergravity action. Useful references are \cite{Green}.

The low-momentum expansion of the IIB superstring leads to type IIB supergravity and a series of higher derivative corrections which can be written as a series in $\alpha'$, the fundamental string length:
\be
\alpha'^4 S_{IIB}=S^{(0)}+\alpha'S^{(1)}+...+(\alpha')^n S^{(n)}+...
\ee
There are no $n=1$ or $n=2$ terms at tree-level and one-loop in the string coupling, and they are not expected to appear at all so that the first correction to the action is an $\alpha'^3$ effect relative to $S^{(0)}$. Since $[\alpha']=[L^2]$, this correction corresponds to terms with eight derivatives. There are ambiguities in these terms, since string amplitudes only determine the action up to terms which vanish on-shell. In a certain scheme one can write the well-known $R^4$ term \cite{GrossWitten} in terms of the Weyl tensor $C$:
\bea
&&\frac{c_1}{\alpha'}\int \ud^{10}x \sqrt{-g} e^{-\phi/2} f^{(0,0)}(\tau,\bar \tau)C^4 \label{C4}\\
C^4&=& -\frac 14 C_{pqrs}C_{pq}^{\ \ tu}C_{rt}^{\ \ vw}C_{suvw}
             +C^{pqrs}C_{p\ r}^{\ t \ u} C_{t\ q}^{\ v \ w}C_{uvsw} \nonumber.
\eea
Here $c_1$ is a constant and $\tau=\tau_1+i \tau_2=C^{(0)}+i e^{-\phi}$ is the complex scalar field, where $C^{(0)}$ is the Ramond-Ramond scalar and $\phi$ is the dilaton. The field $\tau$ parameterises the coset space $SL(2,R)/U(1)$. The function $f^{(0,0)}(\tau,\bar \tau)$ is given by the Eisenstein series
\be
f^{(0,0)}(\tau,\bar \tau)= \sum_{(m,n)\neq (0,0)} \frac{\tau_2^{3/2}}{|m+n \tau|^{3/2}}.
\ee
The exact form of this correction was shown to be a consequence of full non-linear supersymmetry in \cite{Sethi}. The idea is to impose closure of the on-shell supersymmetry algebra order by order in $\alpha'$, which can be used to determine the modular form $f^{(0,0)}$.

There are many terms that are related to $C^4$ by supersymmetry. Among these, we are particularly interested in the ones involving only the five-form $F_5$ and the metric. There is a large class of solutions where these are the only relevant fields, such as the superstar geometries \cite{Superstar1}\cite{Superstar2}, the bubbling solutions of Lin, Lunin and Maldacena \cite{LLM} and the Gutowski-Reall black holes \cite{Gutowski}. Knowing these companion terms to $C^4$ would allow us, in particular, to check under which conditions these solutions receive corrections at $O(\alpha'^{-1})$. The supersymmetric completion of the $C^4$ term when only the five-form is present was suggested in \cite{Green}, and we review the argument next.

The physical content of Type IIB supergravity can be packaged in a scalar superfield $\Phi(x,\theta)$, where $\theta^a, (a=1,...,16)$ is a complex Weyl spinor of $SO(1,9)$. The superfield obeys the conditions
\be
\bar D \Phi=0, \qquad \bar D^4 \bar \Phi=0=D^4 \Phi
\ee
where the first constraint insures independence of $\bar \theta$, and the last two inforce the free field equations of motion on the components of $\Phi$. We can write $\Phi$ as
\bea
\Phi=\tau+\theta \Lambda+\theta^2(G+...)+\theta^3(\mathcal D \psi+...)+\nonumber \\
\theta^4(R+DF+FF+...)+\theta^5(\mathcal D\mathcal D \bar \psi+...)+...+\theta^8(D^4 \bar \tau+...)
\eea
where the dots represent the terms that make each expression supercovariant. Under a supersymmetric transformation labeled by $\epsilon$, we have $\delta_{\epsilon} \Phi=\epsilon \partial \Phi/\partial \theta$. Since the supersymmetry transformations are well known \cite{SugraIIB} we can use them to determine the exact form of the components of the scalar superfield.
In particular,
\be
\delta_{\epsilon} \psi_M=(D_M+\frac{i}{16\cdot 5!} \Gamma^{N_1...N_5}F_{N_1...N_5}\Gamma_M)\epsilon+...\equiv \mathcal D \epsilon
\ee
which implies that
\be
\delta_{\epsilon}(\mathcal D_{[M} \psi{_N]}+...)=(\mathcal R_{MN}+...)\epsilon \label{Dpsi},
\ee
where
\bea
\mathcal R_{MN}=\frac 18 R_{MNPQ}\Gamma^{PQ}-\frac{i}{16\cdot 5!} \Gamma^{K_1...K_5}\Gamma_{[M}D_{N]}F_{K_1...K_5} \nonumber \\
-\frac 1{(16\cdot 5!)^2} \Gamma^{K_1...K_5}\Gamma_{[M}\Gamma^{L_1...L_5}\Gamma_{N]}F_{K_1...K_5}F_{L_1...L_5}.
\eea

Therefore, we determine the quartic term of the scalar superfield to be
\be
(\theta \Gamma^{MNP}\theta) \theta \Gamma_P \mathcal R_{MN}\theta=(\theta \Gamma^{MNP}\theta) (\theta \Gamma^{QRS}\theta)\mathcal R_{MNPQRS}
\ee
where
\bea
\mathcal R_{MNPQRS}&=& \frac 18 g_{PS} R_{MNQR}+\frac i{48} D_M F_{NPQRS}\nonumber \\
&+& \frac 1{384} F_{MNPTU}F_{QRS}^{\ \ \ \ TU}\label{DefR}.
\eea
This expression should be appropriately symmetrized as implied by the contraction with the gamma matrices.

One can use linearized supersymmetry to show that the interactions in $S^{(3)}$ are contained in the integral of a function of $\Phi(x,\theta)$ over half the superspace. On the other hand, non-linear supersymmetry shows that the coefficient of the $C^4$ term must be $f^{(0,0)}$. We are then led to the proposal that $C$ and $F_5$ are present in $S^{(3)}$ in the combination
\bea
S^{(3)}_{\mathcal R^4}&=& \int \ud\! ^{10} x \sqrt{-g} f^{(0,0)}(\tau,\bar \tau) I_{\mathcal R^4} \nonumber \\
I_{\mathcal R^4}&=& \int \ud\! ^{16}\theta[(\theta \Gamma^{MNP}\theta)( \theta \Gamma^{QRS}\theta) \mathcal R_{MNPQRS}]^4+c.c. \label{R4}.
\eea
where the result is written in the Einstein frame. Besides the $C^4$ term, there are other contributions with well defined relative coefficients, such as $F_5^8$, $(\nabla F_5)^4$ and various cross terms. Notice this is the same result obtained more rigourously in \cite{Rajaraman}.

\subsection{Representation content of the integral}

The Grassmannian integral (\ref{R4}) looks rather formidable. Direct evaluation on a specific background would involve doing a summation of $16!>10^{13}$ terms, and it still wouldn't tell us the tensorial form of the corrections to the equations of motion.
We want to compute the tensor structure of the terms packaged in the integral $I_{\mathcal R^4}$. We start by writing
\be
I_{\mathcal R^4}=I^{i_1 j_1 k_1 ... i_8 j_8 k_8} \mathcal R_{i_1 j_1 k_1 i_1 j_2 k_2}... \mathcal R_{i_7 j_7 k_7 i_8 j_8 k_8} \ee
where
\be
I^{i_1 j_1 k_1 ... i_8 j_8 k_8}=\int \ud^{16} \theta (\bar \theta \Gamma^{i_1 j_1 k_1} \theta)...(\bar \theta \Gamma^{i_2 j_2 k_2} \theta). \label{I8}
\ee

This is the symmetric product of eight three-indexed antisymmetric tensors. Group theory tells us that this product contains 33 scalars in ten dimensions, and 24 in eleven dimensions. We conclude that there are 24 parity-even and 9 parity-odd Lorentz singlets in this integral. Using a graph-based approach, a particular basis for these singlets was found in \cite{Superfield}, along with their respective weights in the integral (\ref{I8}). These results are presented in tables \ref{Even},\ref{Odd} in the Appendix. Thus one may recast the integral $I_{\mathcal R^4}$ as a sum of Lorentz scalars:
\be
I_{\mathcal R^4} =\left (\sum_i a_i s_i+\sum_j b_j e_j \right)^{i_1 j_1 k_1 ... i_8 j_8 k_8} \mathcal R_{i_1 j_1 k_1 i_1 j_2 k_2}... \mathcal R_{i_7 j_7 k_7 i_8 j_8 k_8}. \label{SumIR4}.
\ee
In this way the tensorial structure of $I_{\mathcal R^4}$ is made manifest, and amenable to calculation.

Now let us consider a single factor of
\be
(\theta \Gamma^{MNP}\theta)( \theta \Gamma^{QRS} \theta) \mathcal R_{MNPQRS}.
\ee
Necessarily the indices $MNP$ and $QRS$ will be antisymmetrized, and only the part of $\mathcal R$ that is symmetric under the interchange of these triplets will be relevant. However, there are further restrictions coming from various Fierz identities. In fact we have that
\be
(\mbf{16}\otimes \mbf{16}\otimes \mbf{16}\otimes \mbf{16})=\mbf{770}\oplus \mbf{1050^+},
\ee
and so the $SO(9,1)$ representation content of $\mathcal R$ is reduced to $\mathbf{770}+\mathbf{1050^+}$. In particular, this implies that in equation (\ref{DefR}) only the Weyl part of the Riemann tensor is important (this is the $\mathbf{770}$). The $\nabla F_5$ term only contains a $\mathbf{1050^+}$ representation, and this means that in practice only its traceless self-dual part will be relevant, that is we can impose from the start
\be
\nabla_{a} F^{a}_{\ b c d e}=0, \qquad F_5=\star F_5. \label{F5}
\ee
The $(F_5)^2$ term does not contain a $\mathbf{770}$ piece. Its $\mbf {1050^+}$ content is given by applying the relevant Young projector,
\bea
(T_{a b c,d e f})_{|1050^+}&=& \frac 12\left [\frac 12 (T_{a b c, d e f}-3 T_{a b f, d e c}-T_{p a b, d e p}\delta_{f c}+2T_{p a e, p d b}\delta_{f c})\right. \nonumber \\
&-& \left. \frac 1{4!} \epsilon_{a b c d e}^{\ \ \ \ \ p_1...p_5} T_{p_1...p_5 f}\right ].\label{P1050}
\eea
where antisymmetrization in each triplet $[a,b,c]$, $[d,e,f]$ is implied, as well as symmetrization in the pair of triplets. If we impose self-duality of the five-form this projector reduces to
\be
(F_{a b c m n}F_{d e f}^{\ \ \ m n})_{|1050^+}=\frac 12(F_{a b c m n}F_{d e f}^{\ \ \ m n}-3 F_{a b f m n}F_{d e c}^{\ \ \ m n}) \label{F2}
\ee
where once again the right-hand side should be antisymmetrized appropriately. The tensor $\nabla F_5$ is already in the right representation if it obeys the lowest order equations of motion.

The previous equations determine the parts of the five-form that contribute to the $\alpha'$ corrections. 
Similarly, only the Weyl part of the Riemann tensor comes into $I_{\mathcal R^4}$. These conditions will be important later, since if we impose them from the start we can simplify matters tremendously. On the other hand they also determine in which cases the five-form can be neglected or not, which up to now had been dealt with in an {\it ad hoc} fashion in the literature.

\label{SUSY_Corrections}
\subsection{Corrections to supersymmetric solutions}

For solutions of the type IIB action involving only a non-constant metric and five-form, and that in addition preserve some fraction of the supersymmetry, one can already make some statements regarding $\alpha'$ corrections \cite{Green}. Set all fields to zero except for the metric and the five-form. Then equation (\ref{Dpsi}) implies
\be
[\mathcal D_M,\mathcal D_N]\lambda=\mathcal R_{MN} \lambda
\ee
for any spinor $\lambda$. Now suppose the background preserves some fraction of the supersymmetries. Then there is a non-trivial solution for the Killing spinor $^0\psi$ depending on a number of free parameters that corresponds to the number of preserved supersymmetries. Clearly the left-hand side of the equation above vanishes when applied on a Killing spinor, so we get
\be
[\mathcal D_M,\mathcal D_N] ^0\psi=0=\mathcal R_{MN}\ ^0\psi,
\ee
which we can rewrite as
\be
(\theta \Gamma^{MNP}\theta)( \theta \Gamma^{QRS} \ ^0\psi) \mathcal R_{MNPQRS}=0. \label{KilCondition}
\ee
This is nothing but the usual integrability condition on the Killing spinor, which tells us which projection conditions it satisfies. Now consider the object
\be
\mathcal R_{\alpha \beta \gamma \delta}\equiv \Gamma^{MNP}_{[\alpha \beta}\Gamma^{QRS}_{\gamma \delta]}\mathcal R_{MNPQRS},
\ee
through which $I_{\mathcal R^4}$ can be written as
\be
I_{\mathcal R^4}=\int d^{16}\theta ~(\theta^{\alpha}\theta^{\beta}\theta^{\gamma}\theta^{\delta}\mathcal R_{\alpha \beta \gamma \delta})^4=(\mathcal R^4)_{[\alpha_1 \alpha_2 ... \alpha_{15} \alpha_{16}]}.
\ee 
If there is some value of the spinor index $\alpha$ such that $\mathcal R_{\alpha \beta \gamma \delta}$ vanishes, then clearly so does $I_{\mathcal R^4}$. The condition (\ref{KilCondition}) tells us that, in an appropriate basis, there are precisely $\mathcal N$ such values, where $\mathcal N$ is the number of preserved supersymmetries of the background. Since the dilaton multiplies $\mathcal R^4$, this leads us to conclude that, for supergravity solutions involving only the metric and the five-form that preserve at least one supersymmetry, the dilaton will not get sourced by these terms. 

The corrections to the equation of motion for the metric and the five-form are obtained by considering
\be
\delta S^{(3)} \propto \int \ud^{16} \theta~ [(\theta \Gamma^{MNP}\theta)( \theta \Gamma^{QRS}\theta) \mathcal R_{MNPQRS}]^3 \delta \mathcal R=(\mathcal R^3 \delta \mathcal R)_{[\alpha_1 \alpha_2 ... \alpha_{15} \alpha_{16}]}.
\ee
If a solution preserves more than four supersymmetries, then necessarily the $\mathcal R^3$ factor will vanish since each factor of $\mathcal R$ annihilates the Killing spinor. We conclude that
\begin{center}
{\it If a solution is more than $1/4$ BPS, then it receives no corrections at $O(\alpha'^{-1})$.}
\end{center}

In particular, $1/2$ BPS solutions like the LLM bubbling geometries \cite{LLM} do not receive corrections at $O(\alpha'^{-1})$. The full D3-brane solution has been shown explicitly to remain unrenormalized to this order \cite{Green}.

\section{Computing $I_{\mathcal R^4}$}

\subsection{Outline of the computation}

The computation of the $O(\alpha'^{-1})$ corrections reduces to performing the tensor contractions in equation (\ref{SumIR4}). However there is still a long way to go before getting an explicit, tractable result. First of all, there are still too many terms. For instance, each of the parity-even terms in table \ref{Even} has to be properly symmetrized, leading to $(3!)^8\times 8!$ terms. One can take advantage of the fact that the singlets are contracted into the fourth power of $\mathcal R$, and symmetrize this instead. Imposing $\mathcal R=\mathcal R_{[i_1 j_1 k_1] [i_2 j_2 k_2]}$ and symmetry for the interchange $1-2$, then $\mathcal R_{i_1 j_1 k_1 i_2 j_2 k_2}...\mathcal R_{i_7 j_7 k_7 i_8 j_8 k_8}$ can be symmetrized with only hundred and five terms corresponding to various permutations of $1,...,8$. This combination is then contracted against the unsymmetrized Lorentz singlets $s_i$ and $e_i$, leading to a set of scalars in $\mathcal R^4$.

The resulting sum of monomials in $\mathcal R^4$ is very special, because it should be explicitly zero if evaluated on an $\mathcal R$ which is not in the $\mbf {1050^+}$ or $\mbf {770}$ irreps. For this to be true, it can only be that the sum of terms is explicitly Young projected. That is, if we substitute each $\mathcal R$ in the sum by its Young projection into these irreps, then after simplification we should end up exactly with the same set of terms. This is a very strong constraint on the sum, and is a useful check on our computation. It also means that one can choose to work with a tensor $\mathcal R$ which is explicitly in these representations from the start. In particular, it is useful to perform the substitution:
\be
\mathcal R_{a b c d e f}\to \frac 18 g_{a d}C_{b c e f}+\frac 1{48} \mathcal T_{a b c d e f} \label{Substitution}
\ee
where $\mathcal T_{a b c d e f}$ is the piece of $\mathcal R$ in the $\mbf {1050^+}$ irrep, namely
\be
\mathcal T_{a b c d e f}=P_{\mbf{1050^+}}\left (i \nabla_{a} F_{b c d e f}+\frac 18 F_{a b c m n}F_{d e f}^{\ \ \ m n}\right) \label{defineT}
\ee
and $C$ is the Weyl tensor. The $g~\!C$ piece should be appropriately symmetrized. This form for $\mathcal R$ is useful since it allows us to simplify all terms in the sum that have traces. In particular, all double traces of $\mathcal R$ vanish.

One can then focus on the five different types of resulting terms separately, namely $C^4$, $C^3 \mathcal T$, and so forth all the way to $\mathcal T^4$. To simplify each set of terms one must use the various symmetries of the tensors. These include not only mono-term symmetries like antisymmetry or symmetry of indices, but also multi-term symmetries like the cyclic Ricci identity for the Weyl tensor. The former involves defining a canonical order for the index structure, a prescription on how we should rearrange the indices in an expression so that it can be compared to others. It was only a few years ago \cite{Portugal} that an efficient algorithm was designed that accomplishes this canonicalisation for arbitrary symmetries. The multi-term symmetries are encoded in the Young projector of a tensor after modding out by monoterm symmetries. The resulting expressions have the multi-term symmetries explicit, so one does not need to impose them. Using these symmetries one can build basis of independent monomials, in terms of which all terms can be written, leading to compact expressions.

There is a set of identities which can be used to further reduce the number of independent scalars. These are dimensionally dependent identities (also known as Schouten or Lovelock identities) which arise by antisymmetrizing over $d+1$ indices in $d$ dimensions. The $C \mathcal T^3$ and $\mathcal T^4$ scalars have at least twenty-two indices, which allows one to antisymmetrize over eleven of them and contract with the remaining, leading to complicated relations between scalars. We explicitly build these identities for $C \mathcal T^3$ type scalars.

One can avoid working with the parity-odd singlets $e_i$ by imposing $\mathcal T$ to obey a self-duality condition. This is analogous to throwing away all terms involving double traces of $\mathcal R$ and using the Weyl tensor instead of the Riemann curvature. The construction of a basis for monomials that include the tensor $\mathcal T$ is made difficult by the fact that the Young projector for $\mathcal T$ includes an epsilon tensor. We split the problem into two steps, obtaining first a basis of monomials in which $\mathcal T$ is taken to be in the $\mbf {2100}$ representation and later finding relations between the basis elements when self-duality is imposed. It turns out that these relations already include dimensionally dependent identities. This leads to a new method of determining these identities, previously unknown to the author's knowledge.

Once one has constructed monomial basis, the results can be written in terms of these by solving a linear system of equations. Along the way several consistency checks on the result were performed.

\subsection{The computer packages}

The computations described above involve the manipulation of typically thousands of terms.  The main package that was used is the recently released Cadabra \cite{Cadabra}. Among many other functionalities it allows the definition of tensors directly through their Young tableaux and includes the \cite{Portugal} algorithms for canonicalisation. The usage of Young tableaux allows for the program to recognize multi-term symmetries. In particular, one can compute basis of tensor monomials like $R^4$, or decompose a set of terms into one. All tensor manipulation and simplification were performed in Cadabra. The computation of basis of tensors which satisfy a self-dual property is not yet implemented in this package, which led to the use of Mathematica \cite{Mathematica} for basis building and decomposition for all scalars that involved the tensor $\mathcal T$.

\subsection{$\mathcal R^4$ terms and parity matters}

As a first step we need to contract the $24+5$ Lorentz singlets of tables \ref{Even}, \ref{Odd} with the 105 terms coming from symmetrizing the tensor
\be
\mathcal R_{i_1 j_1 k_1 i_2 j_2 k_2}... \mathcal R_{i_7 j_7 k_7 i_8 j_8 k_8}
\ee
over 1,...,8. The resulting contractions are simplified by canonicalisation and by not including any terms which involve double traces of $\mathcal R$. This is justified since we are going to substitute $\mathcal R$ by terms which have no double traces. The parity-even terms follow straightforwardly by this procedure, giving rise to over 450 monomials which we will indicate schematically as $\sum_{even} \mathcal R^4$. The parity-odd ones are much harder, since they involve a ten-dimensional epsilon tensor, leading to tensor objects sporting over thirty indices. The resulting expressions are very hard to canonicalise. Luckily, one does not need to include these parity-odd terms at all.

The specific sum of Lorentz singlets given by tables \ref{Even}, \ref{Odd} in the Appendix has necessarily to be very special. Consider the following simple example. Take $G_5$ to be a 5-form in ten dimensions, and $\hat G_5=\frac 12 (1+\star)G_5$ to be its self-dual part. Then we have
\be
\frac 12 G_{a b c d e}G^{a b c d e}+\frac 1{2.5!}\epsilon^{a b c d e f g h i j}G_{a b c d e}G_{f g h i j}=\hat G_{a b c d e} \hat G^{a b c d e}.
\ee
On the LH side we have a specific combination of a parity-even and parity-odd contractions of $G_5$, which can be written as a single scalar that involves only $\hat G_5$. Analogously, the sum of parity-even and parity-odd singlets involved in the integral (\ref{I8}) are such that when contracted on $\mathcal R^4$ the resulting sum of monomials can be rewritten in terms of scalars composed of the sum of the $\mathbf{770}$ and $\mathbf {1050^+}$ pieces of $\mathcal R$. 

The parity-odd monomials are related to the fact that the integral picks out the $\mathbf {1050^+}$ representation instead of the $\mathbf {1050^-}$ one. The sign of these terms reflects the choice. In particular, suppose one imposes the self-duality condition on $\mathcal R$ from the start. Then at the end we have a sum of monomials of the symbolic form
\be
I_{\hat \mathcal R^4} =\sum_{\mbox{p-even}} (\hat \mathcal R^4)+\sum_{\mbox{p-odd}} (\epsilon \hat \mathcal R^4)
\ee
where $\hat \mathcal R$ only contains the $\mbf {1050^+}$ irrep. Then necessarily, flipping the sign of the parity-odd contribution will result in the integral evaluating to zero. We conclude that
\be
\sum_{\mbox{p-even}} (\hat \mathcal R^4)=\sum_{\mbox{p-odd}} (\epsilon \hat \mathcal R^4).
\ee
If we perform the split $\mathcal R\to g C+\mathcal T$ as in (\ref{Substitution}), where $\mathcal T$ is in the $\mbf {1050^+}$ irrep, we will get terms of the form $C^4$, $C^3 \mathcal T$, ..., $\mathcal T^4$. The argument above then generalizes to the statement that the parity-odd singlets contribute exactly the same as the parity-even ones in all terms that involve at least one $\mathcal T$. As long as we impose the self-duality condition by hand, we need not worry about the contribution of the parity-odd terms, since they will give the same result as the parity-even ones. This is explicit in the example with $G_5$ that was given above: imposing self-duality on $G_5$ on the LH side one sees that the parity-odd term contributes exactly the same as the parity-even one. 

In what follows we will only consider the contribution of the parity-even terms. We proceed by performing the substitution (\ref{Substitution}) and studying each set of terms with differing powers of $C$ separately. In the end the contribution of the parity-odd singlets can be obtained by simply doubling the coefficients of all terms with at least one power of $\mathcal T$.

\subsection{$C^4$ terms}

We start by studying the $C^4$ type terms, for which the final answer is already known and given in (\ref{C4}). The first step is to take $\sum_{even} \mathcal R^4$ and perform the substitution
$$
\mathcal R_{a b c}^{d e f} \to \delta_{[a}^{[d} C_{b c]}^{\ \ e f]}. \label{gC}
$$
This has to be done with care, since it corresponds to a $9^4$ increase in the already large
number of terms. The best way is to replace each $\mathcal R$ in turn, contracting the Kronecker delta and simplify by using the symmetry properties of $\mathcal R$ and $C$. At the end one gets only ten terms, but we can further simplify by decomposing them into a basis of $C^4$ monomials. The construction of this basis is performed by Cadabra, and the result is given in the appendix. The terms are decomposed with respect to this basis, giving the answer
\be
\simeq (0,-1/2,0,0,0,1,0). \label{VecC4}
\ee
Explicitly, this is
\be
C^4=-\frac 12 C_{a b c d} C^{a b}_{\ \ e f} C^{c e}_{\ \ g h} C^{d g f h}+C_{a b c d} C^{a \ c}_{\ e \ f} C^{b \ e}_{\ g \ h} C^{e g f h}.\label{C^4}
\ee
This doesn't seem to match the known form (\ref{C4}). Nevertheless one can show that the cyclic Ricci identity for the Weyl tensor implies the equality of the two expressions.

Quite generally, the way to take into account multi-term symmetries \cite{Superfield} is by using Young-projectors. Consider the Riemann tensor $R_{a b c d}$. The Young projector involves $(2!)^4=16$ terms. One can mod out the monoterm symmetries to reduce this to three:
\be
R_{a b c d}=\frac 13(2 R_{a b c d}-R_{a d b c}+R_{a c b d})
\ee
The RH side then explicitly 'knows' about the cyclic Ricci identity, as can be easily checked. In fact this procedure is completely general. All the symmetries of a tensor are encoded by its Young projector. After modding out the monoterm symmetries all that is left is an expression that where the multi-term symmetries of the tensor are explcity.

A group theory computation tells us that in ten dimensions there are exactly seven Lorentz scalars composed of four Weyl tensors. That is, after taking into account both monoterm and multiterm symmetries, there are only seven independent monomials. The most we can expect to simplify an expression is to decompose it in terms of a basis of this sort. In practice one must substitute all terms in an expression and all basis elements by their Young projected expressions (modulo monoterm symmetries) and solve a large linear system of equations. This is exactly the operation performed by Cadabra, leading to the result (\ref{VecC4}).

\subsection{$C^3 \mathcal T$ terms}

Starting with $\sum_{even} \mathcal R^4$ and making the substitution $\mathcal R=\hat \mathcal R+\mathcal T$, we keep only terms with a single power of $\mathcal T$ in the resulting sum. Substituting each $\hat \mathcal R$ in turn by $g C$ as in (\ref{gC}), canonicalising and simplifying in each step, we arrive at a short result containing a handful of terms.
It is possible to simplify this result further by constructing a basis for $C^3 \mathcal T$ monomials. However, in this case we can't use Cadabra to do it because of the reasons pointed out in the outline. 

The trick to build such basis is to notice that the expression we arrive to must be a Young projection. That is, if we take each factor of $C$ and $\mathcal T$ in each monomial and replace them by the Young projectors of the $\mbf{770}$ and $\mbf{2100}$ representations respectively, then simplifying and canonicalising the resulting (large) expression should lead to exactly the same one. This was checked to be indeed the case, which provides a non-trivial test of the calculation. Notice that we did not use the $\mbf{1050^+}$ Young projector, since this would lead to the presence of epsilon tensors which would only cancel with the parity-odd terms that we did not keep explicitly.

To construct a basis we consider each element in the sum $\sum C^3 \mathcal T$ separately, and see how each one is expanded out after Young projection. The resulting expressions can be thought of as the rewriting of each monomial in terms of a larger basis, a basis of scalars which are only independent up to multi-term symmetries. We can call this the raw basis, and we are interested in the refined basis which is obtained from this one by taking into account all multi-term symmetries.

Each term in the sum is rewritten as a vector in the raw basis by Young projection. These vectors can be thought of as forming the columns of a matrix, the matrix that receives a monomial or sum of monomials and returns their Young projected version. Notice that this matrix might turn out not to be square, since the Young projection of a certain scalar yields an expression which doesn't necessarily yield all possible independent scalars up to multi-term symmetries. This is simply the statement that a given vector might not have components along every basis element. In particular the sum might not contain all these possible elements of the raw basis. If this happens, we take these and treat them as if 
they were refined, Young projecting them and obtaining a new set of elements of the raw basis. We can proceed in this fashion until the entire raw basis is obtained, along with their Young projections. 

This matrix will have a certain set of independent columns, which represent the elements of the refined basis. We expect this to be the number of independent scalars which may be formed out of tensors in the $\mbf{770}$ and $\mbf{2100}$ representations. Actually, we get half that number, since the $\mbf{1050^+}$ and $\mbf {1050^-}$ irreps contained in $\mbf{ 2100}$ are only distinguishable through the use of an epsilon tensor, which we do not take into account in this matrix.
It turns out there are only two independent monomials of the form $C^3 \mathcal T$, and the entire set of these terms can be written in terms of only one of these \footnote{For convenience tensor monomials will be written henceforth with all indices lower.}:
$$
\sum C^3 \mathcal T \simeq 1290240~C_{a b c d}C_{a e f g}C_{b f h i}\mathcal T_{c d e g h i}. \label{C^3T}
$$

\subsection{$C^2 \mathcal T^2$ terms}

The computation of these terms proceeds along exactly the same lines as the previous case. One can construct a refined basis of independent monomials in $C^2 \mathcal T^2$. The result of the computation can be written in terms of this basis, and we call this the refined result. The difference from the previous case is that since $\mathcal T$ should be taken to obey a self-duality condition, the monomials that make up the refined basis are not completely independent, and the refined result can be simplified even further. To find the relations between refined basis elements we make the replacement
\be
\mathcal T_{a b c d e f}\to \frac 12 \left( \mathcal T_{a b c d e f}+ \frac 1{4!} \epsilon_{a b c d e}^{\ \ \ \ \ p_1...p_5}\mathcal T_{p_1...p_5 f}\right ).
\ee
In the resulting expressions we set to zero the terms with a single epsilon tensor and double the contribution of the remaining, by the same argument that allows us to only consider the parity even terms. The double epsilon tensors can then be rewritten as generalized Kronecker deltas, which expand out to a large set of terms. Decomposing back into the refined basis, one ends up with a linear map in the form of a square matrix.  This matrix acts as a self-dual projector for monomials in $C^2 \mathcal T^2$, and it has exactly ten eigenvalues one and some zeros. One would expect this since ten is exactly the expected number of scalars contained in $(\mbf{770})^2\otimes (\mbf{ 1050^+})^2$ by group theory arguments. One also expects that the refined result might be written as a sum of one eigenvectors of this matrix, since only the self-dual part of $\mathcal T$ should contribute to $\mathcal I_{\mathcal R^4}$. If this wasn't true, it would not be possible to write the full set of $C^2\mathcal T^2$ as a sum of tensors in the $(\mbf{770})^2\otimes (\mbf{ 1050^+})^2$ representations. The refined result indeed satisfies this condition, providing a consistency check on the computation.

Finally, we can pick a set of monomials of the refined basis which are independent even when self-duality is taken into account. Writing the refined result in terms of these we get:
\bea
\sum C^2 \mathcal T^2\simeq 1814400 && C_{a b c d} C_{a b c e} \mathcal T_{d f g h i j} \mathcal T_{e f h g i j} \nonumber \\
+443520 && C_{a b c d} C_{a b e f} \mathcal T_{c d g h i j} \mathcal T_{e f g h i j} \nonumber \\
-241920 && C_{a b c d} C_{a e c f} \mathcal T_{b e g h i j} \mathcal T_{d f g h i j} \nonumber \\
-241920 && C_{a b c d} C_{a e c f} \mathcal T_{b g h d i j} \mathcal T_{e g h f i j}  \nonumber \\
-7096320 && C_{a b c d} C_{a e f g} \mathcal T_{b c e h i j} \mathcal T_{d f h g i j} \nonumber \\
-1612800 && C_{a b c d} C_{a e f g} \mathcal T_{b c e h i j} \mathcal T_{d h i f g j} \nonumber \\
+6773760 && C_{a b c d} C_{a e f g} \mathcal T_{b c f h i j} \mathcal T_{d e h g i j} \nonumber \\
-5806080 && C_{a b c d} C_{a e f g} \mathcal T_{b c h e i j} \mathcal T_{d f h g i j}.\label{C^2T^2}
\eea

\subsection{$C \mathcal T^3$ terms}

The calculation of these terms suffers from a problem. The terms cubic in $\mathcal T$ coming from $\sum_{even} \mathcal R^4$ are not explicitly Young projected. That is, the expression is not invariant upon replacement of $C$ and $\mathcal T$ by their Young projectors, unlike the $C^4, C^3 \mathcal T$ and $C^2 \mathcal T^2$ cases. One may suspect that this is due to the existence of Schouten identities. These identities are derived by antisymmetrizing over $d+1$ indices in $d$ dimensions. The $C \mathcal T^3$ monomials contain 22 indices, half of which can be antisymmetrized and contracted with the other half, leading to non-trivial expressions which must be set to zero in ten dimensions.

We start by taking the $C \mathcal T^3$ terms and subtracting off their Young projected version. We want to prove the remainder vanishes in ten dimensions. To do this we must find the relevant Schouten identities. One way to do it is to take one of the terms in the remainder and try to construct such an identity out of it. Starting with
\be
C_{a b c d} \mathcal T_{a b e f g h} \mathcal T_{c d e i j k} \mathcal T_{f g h i j k} 
\ee
then an expression which should be zero in ten dimensions and also contain this term is given by
\be
C_{i_1 i_2 c d} \mathcal T_{a b e i_3 i_4 i_5}\mathcal T_{i_6 i_7 i_8 i j k}\mathcal T_{f g h i_9 i_{10} i_{11}},
\ee
antisymmetrized in $1,2,...,11$ and contracted with $g^{i_1 a} g^{i_2 b}...g^{i_{10} j}g^{i_{11} k}$.
The symmetries of the tensors reduce the number of terms involved in the antisymmetrization, making the calculation feasible. In the end over ninety thousand terms collapse to an expression involving less than two hundred, which must be set to zero in ten dimensions. For the moment we are interested in the part of this expression which is not already Young projected, so we subtract from it its Young projected part. It turns out that this first identity is not enough to cancel the offending terms in the $C\mathcal T^3$ calculation, so we need to find another one. We take a term that doesn't appear in the first identity, 
\be
C_{a b c d} \mathcal T_{a e f b g h} \mathcal T_{c i j e f k} \mathcal T_{d i k g h j}.
\ee
An expression that contains this term and is zero in ten dimensions is given by
\be
C_{i_1 i_2 c d} \mathcal T_{a i_3 i_4 b i_5 h} \mathcal T_{i_6 i_7 i_8 e f k} \mathcal T_{i_{10} i i_9 g i_{11} j}
\ee
antisymmetrized in $1,2,...,11$ and multiplied by $g^{i_1 a} g^{i_2 b}...g^{i_{10} d}g^{i_{11} h}$.
This time the symmetries of the tensors only reduce the 11! terms in the antisymmetrization by a factor of $2!~2!~2!~3!$, making the computation very difficult. Nevertheless it is possible to carry it out, and the resulting expression, after being purged of its Young projected piece, combines with the first identity to precisely cancel out the $C \mathcal T^3$ terms that are left after Young projection. We conclude that the set of $C\mathcal T^3$ coming from $\sum_{even}\mathcal R^4$ are explicitly Young projected up to dimensionally dependent identities\footnote{The length of these identities prevents us from showing them here.}, as they should.

\subsubsection{A new method for computing Schouten identities}
After this first consistency check, the calculation is carried out along the same lines as for the $C^2 \mathcal T^2$. A refined basis is computed and it is shown in the appendix. In terms of this basis the $C\mathcal T^3$ terms can be written compactly. This refined result can once again be simplified further by using the self-duality condition on $\mathcal T$ to find relations between the refined basis elements. However, the self-dual projection matrix turns out to be very peculiar, for it is a defective matrix. Further, the refined result is not a one eigenvector of this matrix, as it should be. We once again suspect that the culprit of this inconsistency is a Schouten identity. This suspicion is reinforced by putting the defective matrix in Jordan normal form, where it is constituted by a diagonal $13\times 13$ block of zeroes, a $5\times 5$ diagonal block of ones, and a $2\times 2$ Jordan block of the form
\be
\left( \begin{tabular}{cc}
0 & 1 \\
0 & 0 
\end{tabular}
\right).
\ee
Inspection of the $\mathcal C\mathcal T^3$ refined result reveals that it only has components along the block of ones and the Jordan block, where it looks like $(A,0)$. If this piece of the refined result could be set to zero this would show the correctness of the computation. Taking the previously discovered Schouten identities and Young projecting them, one finds upon decomposition into the refined basis they are equal, and exactly match the monomial sum that corresponds to $(A,0)$. We are then justified in setting this component of the refined result to zero, and so it becomes a one eigenvector of the self-dual projection matrix as we expect. Alternatively, we can use the Schouten identity to eliminate one of the elements of the refined basis. This reduces the dimension of the self-dual projection matrix by one, and also makes it diagonalizable with eigenvalues zero or one. The refined result becomes a one eigenvector of this matrix.

In this way, not only have we shown that our computation passes a very non-trivial test of correctness, but we have also found a new method of computing Schouten identities. Nowhere in the computation of the self-dual projection matrix is it required to perform antisymmetrizations of any sort, yet Schouten identities show up very naturally by looking at its Jordan normal form. Any deviations from the expected pattern of diagonal blocks of ones and zeroes signals the existence of such identities. These deviations show up in two forms:
\begin{itemize}
\item The existence of non-diagonal Jordan blocks. This is the case we've just analysed. Schouten identities correspond to the zero eigenvectors of these blocks.
\item The existence of eigenvalues different from one or zero. This case will occur in the $\mathcal T^4$ terms. In this case Schouten identities are necessarily the eigenvectors corresponding to these eigenvalues.
\end{itemize}

The fact that one can find out about dimensionally dependent identities from looking at self-dual projection matrices is not totally unexpected, since the computation of these brings in necessarily epsilon tensors, which ``know'' about the dimension of the vector space these tensors live in. 

We conclude this subsection by giving the final answer for the $C\mathcal T^3$ terms.  We can pick a subset of refined basis elements that are still independent after self-duality is imposed, and write the refined result in terms of these. We find,
\bea
\sum C\mathcal T^3 \simeq 483840 && C_{a b c d} \mathcal T_{a b e f g h} \mathcal T_{c d e i j k} \mathcal T_{f g h i j k} \nonumber \\
- 4354560 && C_{a b c d} \mathcal T_{a b e f g h} \mathcal T_{c d f i j k} \mathcal T_{e g h i j k} \nonumber \\
- 17418240 && C_{a b c d} \mathcal T_{a b e f g h} \mathcal T_{c d f i j k} \mathcal T_{e g i h j k} \nonumber \\
+ 8709120 && C_{a b c d} \mathcal T_{a b e f g h} \mathcal T_{c e f i j k} \mathcal T_{d g h i j k} \nonumber.\label{CT^3}
\eea

\subsection{$\mathcal T^4$ terms}

After what we've learned in the previous cases, there is no conceptual problem in need of tackling for these terms. The $\mathcal T^4$ terms coming from $\sum_{even} \mathcal R^4$ are not automatically Young projected. We consider this to be the fault of Schouten identities, and proceed by taking only the Young projected part of the sum. The refined basis is then obtained straightforwardly. The construction of the self-dual projection matrix is quite involved this time since there are forty refined basis elements carrying four factors of $\mathcal T$, each of which contributes with an epsilon tensor. These epsilon tensors can pair up in six possible ways as well as all at once. Each pairing contributes a generalized Kronecker delta which breaks up into possibly tens of thousands of terms, making this calculation the most computationally intensive part of this work.

The self-dual projection matrix that results has several eigenvalues which are not one or zero. These can't be solved for explicitly since they are the five real roots of a quintic polinomial, which makes it hard to write down the associated eigenvectors/Schouten identities. The way around this is to take the matrix that equals this polinomial evaluated at the self-dual projection matrix. In this way the space of Schouten identities is mapped onto the null space of this new matrix, and we can build a basis for it. We proceed by eliminating some of refined basis elements by using these Schouten identities. In the end the self-dual projection matrix only has a set of zero eigenvalues plus five one eigenvalues, which corresponds to the number of scalars present in the tensor product of four $\mbf{1050^+}$ irreps. Further the refined result is exactly a one eigenvector of this matrix, providing another non-trivial check on our computation. Finally, we can find a reduced basis and write the refined result in terms of these:
\bea
\sum \mathcal T^4\simeq 5153760 && \mathcal T_{a b c d e f} \mathcal T_{a b c d g h} \mathcal T_{e g i j k l} \mathcal T_{f i j h k l} \\
- 7925040 && \mathcal T_{a b c d e f} \mathcal T_{a b c d g h} \mathcal T_{e i j g k l} \mathcal T_{f i k h j l}  \\
- 2799360 && \mathcal T_{a b c d e f} \mathcal T_{a b c g h i} \mathcal T_{d e j g k l} \mathcal T_{f h k i j l}  \\
+ 22394880 && \mathcal T_{a b c d e f} \mathcal T_{a b c g h i} \mathcal T_{d g j e k l} \mathcal T_{f h k i j l}  \\
+  5806080 && \mathcal T_{a b c d e f} \mathcal T_{a b d e g h} \mathcal T_{c g i j k l} \mathcal T_{f j k h i l}
 \label{T^4}
\eea

\section{Summary of results}

Equations (\ref{C^4}),(\ref{C^3T}),(\ref{C^2T^2}),(\ref{CT^3}) and (\ref{T^4}) constitute the main results of this paper. These terms represent the contribution of the even parity Lorentz singlets to $I_{\mathcal R^4}$. The contribution of the parity-odd terms is given as we've seen by simply doubling the terms which contain at least one $\mathcal T$.  The result can be written as:
\be
\mathcal I_{\mathcal R^4} \propto    \mathcal W  \equiv \frac{1}{86016}\sum_i n_i M_i \nonumber \\ 
\ee
\begin{center}
\begin{tabular}{|c|c|}
\hline $n_i$  & $M_i$ \\ \hline
-43008 & $ C_{a b c d} C_{a b e f} C_{c e g h} C_{d g f h} $ \\ \hline
86016 & $  C_{a b c d} C_{a e c f} C_{b g e h} C_{d g f h}$ \\ \hline
129024 & $ C_{a b c d} C_{a e f g} C_{b f h i} \mathcal T_{c d e g h i}$ \\ \hline
30240 & $ C_{a b c d} C_{a b c e} \mathcal T_{d f g h i j} \mathcal T_{e f h g i j} $ \\ \hline
7392 & $ C_{a b c d} C_{a b e f} \mathcal T_{c d g h i j} \mathcal T_{e f g h i j}$ \\ \hline
-4032 & $ C_{a b c d} C_{a e c f} \mathcal T_{b e g h i j} \mathcal T_{d f g h i j}$ \\ \hline
-4032 & $ C_{a b c d} C_{a e c f} \mathcal T_{b g h d i j} \mathcal T_{e g h f i j}$ \\ \hline
-118272 & $ C_{a b c d} C_{a e f g} \mathcal T_{b c e h i j} \mathcal T_{d f h g i j}$ \\ \hline
-26880 & $ C_{a b c d} C_{a e f g} \mathcal T_{b c e h i j} \mathcal T_{d h i f g j}$ \\ \hline
112896 & $ C_{a b c d} C_{a e f g} \mathcal T_{b c f h i j} \mathcal T_{d e h g i j}$ \\ \hline
-96768 & $ C_{a b c d} C_{a e f g} \mathcal T_{b c h e i j} \mathcal T_{d f h g i j}$ \\ \hline
1344 & $ C_{a b c d} \mathcal T_{a b e f g h} \mathcal T_{c d e i j k} \mathcal T_{f g h i j k}$ \\ \hline
-12096 & $ C_{a b c d} \mathcal T_{a b e f g h} \mathcal T_{c d f i j k} \mathcal T_{e g h i j k} $ \\ \hline
-48384 & $ C_{a b c d} \mathcal T_{a b e f g h} \mathcal T_{c d f i j k} \mathcal T_{e g i h j k}$ \\ \hline
24192 & $ C_{a b c d} \mathcal T_{a b e f g h} \mathcal T_{c e f i j k} \mathcal T_{d g h i j k} $ \\ \hline
2386 & $ \ \mathcal T_{a b c d e f} \mathcal T_{a b c d g h} \mathcal T_{e g i j k l} \mathcal T_{f i j h k l}$ \\ \hline
-3669 & $ \ \mathcal T_{a b c d e f} \mathcal T_{a b c g h i} \mathcal T_{d e j g k l} \mathcal T_{f h k i j l} $ \\ \hline
-1296 & $ \ \mathcal T_{a b c d e f} \mathcal T_{a b c g h i} \mathcal T_{d g j e k l} \mathcal T_{f h j i k l} $ \\ \hline
10368 & $ \ \mathcal T_{a b c d e f} \mathcal T_{a b c g h i} \mathcal T_{d g j e k l} \mathcal T_{f h k i j l} $ \\ \hline
2688 & $ \ \mathcal T_{a b c d e f} \mathcal T_{a g h d i j} \mathcal T_{b g k e i l} \mathcal T_{c h k f j l}$ \\ \hline 
\end{tabular}\label{FinalRes}
\end{center}
The tensor $\mathcal T$ is defined by \ref{defineT}. If we impose self-duality of the five-form, using \ref{F2} this reduces to
$$
\mathcal T_{a b c d e f}=i \nabla_{a} F_{b c d e f}+\frac 1{16}\left (F_{a b c m n}F_{d e f}^{\ \ \ m n}-3 F_{a b f m n}F_{d e c}^{\ \ \ m n}\right),
$$
where RHS should be antisymmetrized in the triplets $[abc],[def]$ and symmetrized for their interchange.

This result constitutes the full set of higher derivative corrections that involve the metric and the five-form at $O(\alpha'^{3})$. The usual pair of $C^4$ terms is accompanied by eighteen other terms that give the contribution of the five-form. We have explicitly tested this result by evaluating it on various supersymmetric solutions, having obtained a vanishing result as expected.

The type IIB supergravity action together with its first non-trivial correction in $\alpha'$ is then given by
$$
S_{IIB}=\frac 1{16 \pi G_N} \int \ud^{10}x\left(R-(\partial \phi)^2-\frac 1{4\cdot 5!} F_5^2+\gamma(\phi) \mathcal W\right)
$$
where $G_N\propto \alpha'^4$ and $\gamma(\phi)=\frac 1{16} (\alpha')^3 f^{(0,0)}(\tau,\bar \tau)$. The equation of motion for the five-form as derived from this action is not consistent with the usual self-duality condition. Defining $\hat \gamma=2\cdot 5! \gamma $, this condition is generalized to \cite{RRsector}
\be
(1-\star)\left(F_5-\hat \gamma(\phi)\frac{\delta \mathcal W}{\delta F_5}\right)=0
\ee
and it is then consistent with the equation of motion.

\section{Application to the thermodynamics of black holes}

In general, if a solution to the lowest-order equations of motion contains a non-trivial five-form, then its contribution at $O(\alpha'^3)$ cannot be neglected. Previous work \cite{Gubser}\cite{Buchel} studied the effect of $\alpha'$ corrections to various asymptotically $AdS_5\times S_5$ black holes. In $\cite{Gubser}$ the authors considered the near-horizon limit of the black D3-brane, which is dual to $N=4 SYM$ at finite temperature. The corrections to the geometry were obtained assuming that the only relevant terms were $C^4$, and from that the first term in the strong coupling expansion of the free energy was obtained. We are now in a position to justify this procedure. An expanded version of these examples and others will be presented in a separate publication

Since we are doing an expansion in powers of $\alpha'$, the corrections to the equations of motion from the $\alpha'$ term in the action are to be evaluated on the lowest order metric and five-form. For the $AdS_5$ black hole solution the five-form is particularly simple, and the tensor $\mathcal T_{a b c d e f}$ is vanishing. This means that all terms quadratic in $\mathcal T$ will vanish upon variation. However, nothing prevents the $C^3 \mathcal T$ term from contributing to the equations of motion. What saves the day here is that the tensor $C^3$ that is contracting $\mathcal T$ must necessarily be in the $\mbf{1050^-}$ irrep to form a Lorentz scalar, and for the $AdS_5$ black hole solution explicit calculation of this tensor gives zero. If $P^{\pm}$ is the projector onto the $\mbf{1050}^{\pm}$ irrep, then
$$
C^3 \mathcal T=(P^{-}C^3) (P^{+}\mathcal T)
$$
and both factors are separately zero on this background.

In general, solutions which have a non-trivial five-form present will receive corrections from the new terms found in this paper. As an application, we will now discuss corrections to the thermodynamics of charged black holes with spherical or flat horizons. The solutions we are intested in are given by
\bea
ds^2_{10}&=& \sqrt{\Delta} \left(-(H_1 H_2 H_3)^{-1}f \ud t^2+(f^{-1} \ud r^2+r^2 \ud \Omega_{(3,k)}^2)\right)\nonumber \\
&+& \frac{1}{\sqrt{ \Delta}}\sum_{i=1}^3 H_i\left(L^2 \ud \mu_i^2+ \mu_i^2(L\ud \phi_i+A_i)^2\right)\nonumber \\
\Delta&=& H_1 H_2 H_3 \sum_{i=1}^3 \frac{\mu_i^2}{H_i}  \qquad H_i= 1+\frac{q_i}{r^2}\nonumber \\
f&=& k-\frac{\mu}{r^2}+\frac{r^2}{L^2}H_1 H_2 H_3 \qquad  A_i=\frac{\tilde q_i}{q_i} (H_i^{-1}-1) \ud t. \label{Metric}
\eea
with $q_i$ related to the physical U(1) charges $\tilde q_i=\sqrt{q_i(k q_i+\mu)}$. The constant $L$ sets the length scale. Here $\Omega_{(3,k)}$ is the 3-manifold of curvature $k=(0,1)$, namely $R^3,S^3$ and the coordinates $\mu_i$, $i=1,2,3$ are constrained by $\mu_1^2+\mu_2^2+\mu_3^2=1$.
The dilaton is constant and the five-form is given by
\be
F_5=\ud B_4+\star ~(\ud B_4), \qquad B_4=-\frac{r^4}L \Delta \ud t \wedge \ud \Omega_{(3,k)}-L \sum_{i=1}^3 \tilde q_i \mu_i^2(L\ud \phi_i-\frac{q_i}{\tilde q_i}\ud t)\wedge \ud \Omega_{(3,k)}.
\ee
The thermodynamics of these geometries have been considered previously in the literature \cite{Chamblin:1999tk}\cite{Cvetic:1999ne}\cite{Buchel:2003re}\cite{Liu:2004it}. In particular \cite{Buchel} considered $\alpha'$ modifications to the thermodynamics of these geometries by computing the corrections to the geometry induced by the $C^4$ term. However, there is no reason to believe the five-form doesn't contribute here, and in fact the full $\mathcal R^4$ set of terms yields quite a different result from simply considering $C^4$. In fact, a direct evaluation  on the above solutions yields
$$
\mathcal W=180 \frac{\mu^4}{x^8} \label{R4SuperStar}
$$
where $x\equiv r^2+Q$, whereas evaluating $C^4$ alone yields a complicated expression. Notice also that this result is very similar to the one obtained for the non-extremal AdS solution \cite{Gubser}
$$
C^4=180 \frac{\mu^4}{r^{16}}.
$$
This lends further credence to the correction of our result. The evaluation of the full $\alpha'$ corrected geometry will not be done here. However, in \cite{Gubser} it was found that a naive computation of the correction to the free energy by evaluation of the $\alpha'$ correction to the action on the lowest-order solution yielded the same answer as a full-fledged calculation including corrections to the geometry. This gives at least some hope that something similar will happen for the charged solution, and in what follows we assume this. We will leave the full computation including corrections to the geometry to future work.

We consider the case where all three charges are the same. The thermodynamical quantity of interest here is the Gibbs free energy density $\Omega=E-TS-\sum_{i} \mu_i \tilde q_i$. Its computation in supergravity corresponds to the strong coupling limit of the free energy of $\mathcal N=4$ SYM in the presence of chemical potentials for the $R$-charges.
To lowest order in $\alpha'$ one has \cite{Buchel}:
$$
\Omega=\frac{N^2}{8\pi^2}\left(-\mu+\frac 34 k^2+2r_+^2k-2 q k \right)
$$
where we have set $L=1$ and $r_+$ is the position of the horizon determined by $f(r_+)=0$. The leading order $\alpha'$ correction is then
\be
\beta \delta \Omega=\delta S^{(3)}=-\beta \frac{\pi^3}{16\pi G_{10}}\int_{r_+}^{+\infty} \ud r~ r(r^2+Q) \gamma \left( \frac{180 \mu^4}{(r^2+Q)^8}\right).
\ee
The AdS/CFT correspondence \cite{Magoo} gives
$$
\frac{1}{16\pi G_{10}}=\frac 1{2\kappa^2}, \qquad L^4=1=\frac{N \kappa}{2\pi^{5/2}}
$$
so we get
$$
\delta \Omega=-15 \frac{N^2}{8\pi^2}\gamma \frac{\mu^4}{x_+^6}.
$$
The $\alpha'^3$ corrected Gibbs free energy is then:
$$
\Omega=\frac{N^2}{8\pi^2}\left(-\mu+\frac 34 k+2 r^+k-2q k-\frac{15}8 \frac{\zeta(3)}{\lambda^{3/2}} \frac{\mu^4}{x_+^6}\right ).
$$
The dependence on temperature is hidden by the relation
$$
2\pi T=\frac{\mu}{x_+^{3/2}}+x_+^{1/2}\left(\frac{x_+-3Q}{x_+-Q}\right)
$$
which can be obtained in the usual fashion from the Euclidean version of the geometry (\ref{Metric}).

We have seen that the exact expression for the ${\alpha'}^3$ corrections that include the RR five-form are of interest in a class of problems in which the three-form strengths vanish and the dilaton-axion is constant.  It would, of course, be very interesting to determine the complete set of higher derivative corrections at this order, but for the moment this seems to be a daunting task. The set of corrections computed in this paper leads to many possible applications. One such application is to verify results existing in the literature that use AdS/CFT methods to compute hydrodynamic coefficients for strongly coupled gauge theories. It would be interesting to see if our results modify existing computations which have only considered $C^4$ corrections \cite{Buchel:2008wy}\cite{Benincasa:2005qc}.

\acknowledgments

The author would like to thank Michael B. Green and Aninda Sinha for getting him interested in this problem and useful discussions. Special thanks go to Kasper Peeters for help with technical topics. This work was supported by the Portuguese government, FCT grant SFRH/BD/23438/2005.

\appendix
\section{Appendix}
\subsection{Decomposition of $I^{i_1 j_1 k_1...i_8 j_8 k_8}$}
\TABLE{
\begin{tabular}{c|c}
Singlet $s_i$ & Coefficient $a_i/(2^{19}\times 3^6)$ \\ \hline
$\delta_{i_1 i_2}\delta_{i_3 i_4}\delta_{i_5 i_6} \delta_{i_7 i_8} \delta_{j_1 j_2}\delta_{j_3 j_4}\delta_{j_5 j_6}\delta_{j_7 j_8}\delta_{k_1 k_2}\delta_{k_3 k_4}\delta_{k_5 k_6}\delta_{k_7 k_8}$ & -269 \\
$\delta_{i_1 i_2}\delta_{i_3 i_4}\delta_{i_5 i_6} \delta_{i_7 j_6} \delta_{i_8 k_5}\delta_{j_1 j_2}\delta_{j_3 j_4}\delta_{j_5 j_7}\delta_{j_8 k_6}\delta_{k_1 k_2}\delta_{k_3 k_4}\delta_{k_7 k_8}$ & 4968 \\
$\delta_{i_1 i_2}\delta_{i_3 i_4}\delta_{i_5 i_6} \delta_{i_7 k_6} \delta_{i_8 k_5}\delta_{j_1 j_2}\delta_{j_3 j_4}\delta_{j_5 j_6}\delta_{j_7 j_8}\delta_{k_1 k_2}\delta_{k_3 k_4}\delta_{k_7 k_8}$ & 7956 \\
$\delta_{i_1 i_2}\delta_{i_3 i_4}\delta_{i_5 j_4} \delta_{i_6 k_3} \delta_{i_7 j_3}\delta_{i_8 k_5}\delta_{j_1 j_2}\delta_{j_5 j_6}\delta_{j_7 k_4}\delta_{j_8 k_6}\delta_{k_1 k_2}\delta_{k_7 k_8}$ & -2304 \\
$\delta_{i_1 i_2}\delta_{i_3 i_4}\delta_{i_5 j_3} \delta_{i_6 k_4} \delta_{i_7 k_3}\delta_{i_8 k_6}\delta_{j_1 j_2}\delta_{j_4 j_5}\delta_{j_6 k_5}\delta_{j_7 j_8}\delta_{k_1 k_2}\delta_{k_7 k_8}$ & 70848 \\
$\delta_{i_1 i_2}\delta_{i_3 i_4}\delta_{i_5 k_3} \delta_{i_6 k_4} \delta_{i_7 k_5}\delta_{i_8 k_6}\delta_{j_1 j_2}\delta_{j_3 j_4}\delta_{j_5 j_6}\delta_{j_7 j_8}\delta_{k_1 k_2}\delta_{k_7 k_8}$ & -24192 \\
$\delta_{i_1 i_2}\delta_{i_3 i_4}\delta_{i_5 k_4} \delta_{i_6 j_5} \delta_{i_7 k_6}\delta_{i_8 j_7}\delta_{j_1 j_2}\delta_{j_3 j_4}\delta_{j_6 k_5}\delta_{j_8 k_7}\delta_{k_1 k_2}\delta_{k_3 k_8}$ & -32544 \\
$\delta_{i_1 i_2}\delta_{i_3 j_2}\delta_{i_4 k_1} \delta_{i_5 i_6} \delta_{i_7 j_6}\delta_{i_8 k_5}\delta_{j_1 j_3}\delta_{j_4 k_2}\delta_{j_5 j_7}\delta_{j_8 k_6}\delta_{k_3 k_4}\delta_{k_7 k_8}$ & -3888 \\
$\delta_{i_1 i_2}\delta_{i_3 j_2}\delta_{i_4 k_1} \delta_{i_5 i_6} \delta_{i_7 k_6}\delta_{i_8 k_5}\delta_{j_1 j_3}\delta_{j_4 k_2}\delta_{j_5 j_6}\delta_{j_7 j_8}\delta_{k_3 k_4}\delta_{k_7 k_8}$ & -26352 \\
$\delta_{i_1 i_2}\delta_{i_3 k_2}\delta_{i_4 k_1} \delta_{i_5 i_6} \delta_{i_7 k_6}\delta_{i_8 k_5}\delta_{j_1 j_2}\delta_{j_3 j_4}\delta_{j_5 j_6}\delta_{j_7 j_8}\delta_{k_3 k_4}\delta_{k_7 k_8}$ & -20412 \\
$\delta_{i_1 i_2}\delta_{i_3 j_1}\delta_{i_4 k_3} \delta_{i_5 k_1} \delta_{i_6 k_4}\delta_{i_7 k_2}\delta_{i_8 k_5}\delta_{j_2 j_3}\delta_{j_4 j_5}\delta_{j_6 j_8}\delta_{j_7 k_6}\delta_{k_7 k_8}$ & 124416 \\
$\delta_{i_1 i_2}\delta_{i_3 j_1}\delta_{i_4 k_2} \delta_{i_5 k_1} \delta_{i_6 j_5}\delta_{i_7 k_5}\delta_{i_8 k_4}\delta_{j_2 j_3}\delta_{j_4 k_3}\delta_{j_6 j_7}\delta_{j_8 k_6}\delta_{k_7 k_8}$ & 10368 \\
$\delta_{i_1 i_2}\delta_{i_3 j_1}\delta_{i_4 k_1} \delta_{i_5 j_2} \delta_{i_6 k_4}\delta_{i_7 k_3}\delta_{i_8 k_2}\delta_{j_3 j_6}\delta_{j_4 j_5}\delta_{j_7 k_5}\delta_{j_8 k_6}\delta_{k_7 k_8}$ & 196992 \\
$\delta_{i_1 i_2}\delta_{i_3 j_2}\delta_{i_4 j_3} \delta_{i_5 k_1} \delta_{i_6 k_2}\delta_{i_7 k_3}\delta_{i_8 k_4}\delta_{j_1 j_4}\delta_{j_5 j_6}\delta_{j_7 k_6}\delta_{j_8 k_7}\delta_{k_5 k_8}$ & -10368 \\
$\delta_{i_1 i_2}\delta_{i_3 j_1}\delta_{i_4 k_3} \delta_{i_5 k_2} \delta_{i_6 k_1}\delta_{i_7 k_6}\delta_{i_8 k_5}\delta_{j_2 j_3}\delta_{j_4 j_5}\delta_{j_6 k_4}\delta_{j_7 j_8}\delta_{k_7 k_8}$ & 373248 \\
$ \delta_{i_1 i_2}\delta_{i_3 j_1}\delta_{i_4 k_3} \delta_{i_5 j_4} \delta_{i_6 k_5}\delta_{i_7 k_2}\delta_{i_8 k_1}\delta_{j_2 j_3}\delta_{j_5 k_4}\delta_{j_6 j_7}\delta_{j_8 k_6}\delta_{k_7 k_8}$ & -331776 \\
$ \delta_{i_1 i_2}\delta_{i_3 j_1}\delta_{i_4 k_2} \delta_{i_5 k_4} \delta_{i_6 k_1}\delta_{i_7 k_5}\delta_{i_8 k_6}\delta_{j_2 j_3}\delta_{j_4 k_3}\delta_{j_5 j_6}\delta_{j_7 j_8}\delta_{k_7 k_8}$ & -165888 \\
$ \delta_{i_1 i_2}\delta_{i_3 j_1}\delta_{i_4 j_2} \delta_{i_5 j_3} \delta_{i_6 k_1}\delta_{i_7 k_4}\delta_{i_8 k_5}\delta_{j_4 k_2}\delta_{j_5 k_3}\delta_{j_6 j_7}\delta_{j_8 k_6}\delta_{k_7 k_8}$ & 41472 \\
$ \delta_{i_1 i_2}\delta_{i_3 k_1}\delta_{i_4 k_2} \delta_{i_5 k_3} \delta_{i_6 k_4}\delta_{i_7 k_5}\delta_{i_8 k_6}\delta_{j_1 j_2}\delta_{j_3 j_4}\delta_{j_5 j_6}\delta_{j_7 j_8}\delta_{k_7 k_8}$ & -10368 \\
$\delta_{i_1 i_2}\delta_{i_3 k_1}\delta_{i_4 k_2} \delta_{i_5 k_3} \delta_{i_6 j_3}\delta_{i_7 k_5}\delta_{i_8 k_6}\delta_{j_1 j_2}\delta_{j_4 j_6}\delta_{j_5 k_4}\delta_{j_7 j_8}\delta_{k_7 k_8}$ & -171072 \\
$\delta_{i_1 i_2}\delta_{i_3 j_1}\delta_{i_4 k_2} \delta_{i_5 k_1} \delta_{i_6 j_5}\delta_{i_7 k_6}\delta_{i_8 k_4}\delta_{j_2 j_3}\delta_{j_4 k_3}\delta_{j_6 k_5}\delta_{j_7 j_8}\delta_{k_7 k_8}$ & -238464 \\
$\delta_{i_1 i_2}\delta_{i_3 k_1}\delta_{i_4 k_2} \delta_{i_5 k_3} \delta_{i_6 k_4}\delta_{i_7 j_5}\delta_{i_8 k_7}\delta_{j_1 j_2}\delta_{j_3 j_4}\delta_{j_6 j_8}\delta_{j_7 k_5}\delta_{k_6 k_8}$ & -248832 \\
$\delta_{i_1 i_2}\delta_{i_3 j_1}\delta_{i_4 k_1} \delta_{i_5 i_8} \delta_{i_6 j_8}\delta_{i_7 k_8}\delta_{j_2 j_5}\delta_{j_3 j_6}\delta_{j_4 j_7}\delta_{k_2 k_5}\delta_{k_3 k_6}\delta_{k_4 k_7}$ & -62208 \\
$\delta_{i_1 i_2}\delta_{i_3 k_2}\delta_{i_4 j_3} \delta_{i_5 k_4} \delta_{i_6 j_5}\delta_{i_7 k_6}\delta_{i_8 j_7}\delta_{j_1 j_2}\delta_{j_4 k_3}\delta_{j_6 k_5}\delta_{j_8 k_7}\delta_{k_1 k_8}$ & 63504 \\ \hline
\end{tabular}
\label{Even}
\caption{Parity-even Lorentz singlets.  Indices [ijk] should be antisymmetrized, and [1-8] symmetrized.}
}

\TABLE{
\begin{tabular}{c|c}
Singlet $e_i$ & Coefficient $b_i/(2^{21}\times 3^6 \times 5)$  \\ \hline
$\delta_{i_1 i_2} \delta_{i_3 i_4} \delta_{i_5 j_3} \delta_{i_6 i_7} \delta_{j_1 j_2} \delta_{j_4 j_5} \delta_{k_1 k_2}\epsilon_{i_8 j_8 k_8 j_7 k_7 j_6 k_6 k_5 k_4 k_3}$& 7 \\ 
$\delta_{i_1 i_2} \delta_{i_3 k_1} \delta_{i_4 k_2} \delta_{i_5 j_3} \delta_{i_6 j_4} \delta_{i_7 j_5} \delta_{j_1 j_2}\epsilon_{i_8 j_8 k_8 j_7 k_7 j_6 k_6 k_5 k_4 k_3}$& 42 \\
$\delta_{i_1 i_2} \delta_{i_3 j_1} \delta_{i_4 j_2} \delta_{i_5 j_3} \delta_{i_6 k_1} \delta_{i_7 k_2} \delta_{j_4 j_5}\epsilon_{i_8 j_8 k_8 j_7 k_7 j_6 k_6 k_5 k_4 k_3}$& -294 \\
$\delta_{i_1 i_2} \delta_{i_3 j_1} \delta_{i_4 i_5} \delta_{i_6 j_4} \delta_{i_7 k_1} \delta_{j_2 j_3} \delta_{j_5 j_6}\epsilon_{i_8 j_8 k_8 j_7 k_7 k_6 k_5 k_4 k_3 k_2}$&  -168\\
$\delta_{i_1 i_2} \delta_{i_3 j_1} \delta_{i_4 j_2} \delta_{i_5 j_3} \delta_{i_6 j_4} \delta_{i_7 j_5} \delta_{j_6 j_7}\epsilon_{i_8 j_8 k_8 k_7 k_6 k_5 k_4 k_3 k_2 k_1}$&  264 \\ \hline 
\end{tabular}
\label{Odd}
\caption{Parity-odd Lorentz singlets. Indices [ijk] should be antisymmetrized, and [1-8] symmetrized. There are four other possible singlets which don't interest us since their contribution to $I_{\mathcal R^4}$ is zero.}
}

\bibliography{mybib}{}
\bibliographystyle{JHEP}

\end{document}